\documentclass{article}

\def\b#1{^{(#1)}}
\def\be{\begin{equation}}
\def\bel{\begin{equation}\label}
\def\ee{\end{equation}}
\def\ba{\begin{array}}
\def\ea{\end{array}}
\def\Journal#1#2#3#4{{ #1}{ \bf #2}, #3 (#4)}
\def\PLB{ {{\it Phys Lett.}} B}

\def\tens{\otimes}
\def\th{\theta}
\def\r#1{(\ref{#1})}
\def\ds{\displaystyle}
\begin{document}

\title{From noncommutative space-time to quantum relativistic symmetries
with fundamental mass parameter}

\author{Jerzy Lukierski \\
\normalsize \sl
\normalsize \sl Institute for Theoretical Physics, University of Wroc\l{}aw,\\
\normalsize \sl pl. Maxa Borna 9, 50-204 Wroc\l{}aw, Poland\\ 
\normalsize \sl E-mail: lukier@ift.uni.wroc.pl
}
\date{December 29, 2001}


\maketitle

\begin{abstract}
{We consider the simplest class of Lie-algebraic deformations of 
space-time algebra, with the selection of $\kappa$-deformations as providing 
quantum deformation of relativistic framework. We recall that the 
$\kappa$-deformation along any direction in Minkowski space can be obtained.
Some problems of the formalism of $\kappa$-deformations will be considered.
We shall comment on the conformal extension of light-like $\kappa$-deformation
as well as on the applications to astrophysical problems.
}
\end{abstract}

\section{Introduction}
Recently the noncommutative modification of space-time geometry gained 
remarkable attention mainly due to two reasons:
\begin{enumerate}
\item[i)] It has been argued (see e.g \cite{1,2,3,4}) that the quantum gravity
corrections can be algebraically described by the noncommutative modification
of Minkowski space-time geometry
\be\label{1-1} 
[x_\mu,x_\nu]=0\quad \longrightarrow \quad [\hat x_\mu,\hat x_\nu]=i\frac1\kappa 
\theta_{\mu\nu}(\frac{\hat x}{\kappa})
\ee
where $\kappa$ is fundamental mass-like parameter and in the general case
\be\label{1-2} 
\hat \theta_{\mu\nu}(\hat x) =\th\b0 _{\mu\nu}+\th\b1_{\mu\nu}{}^\lambda\hat x _\lambda
+\th\b2 _{\mu\nu}{}^{\lambda\tau} \hat x _\lambda \hat x_\tau + \ldots
\ee

\item[ii)] The noncommutativity of space-time coordinates can be obtained also 
from the quantization of open string in $D=10$ in the presence of 
nonvanishing 2-tensor background field $B_{\mu\nu}(x)$ 
(see e.g. \cite{5,6,7}). If $B_{\mu\nu}$  is constant, it implies also constant value 
of $\theta_{\mu\nu}$. In general case the relation \r{1-1} should be treated 
as a part of deformed relativistic phase space 
algebra with generators
$(\hat x_\mu,\hat p_\mu)$ 
and local function $\theta_{\mu\nu}(\hat x)$ extended to the functional 
$\theta_{\mu\nu}[\hat x,\hat p]$ of eight phase space generators.
\end{enumerate}

The most popular case of deformation \r{1-1} is provided by constant 
value of the commutator $(\theta_{\mu\nu}(\hat x)= \theta\b0_{\mu\nu})$. 
Such assumption is also made in original Doplicher-Fredenhagen-Roberts 
(DFR) model \cite{1}. In such a case the relativistic symmetries 
remain classical, however the presence of constant two-tensor 
$\theta\b0_{\mu\nu}$ implies the breaking of classical Lorentz 
symmetry\footnote{It should be mentioned that one can avoid Lorentz symmetry 
breaking in quantum theory, by assuming that $\theta\b0_{\mu\nu}$ is a
constant tensor operator transforming as relativistic two-tensor under 
unitary rotations of Lorentz group in extended Hilbert space.}.
In local QFT models on deformed Minkowski space with constant value of 
$\theta_{\mu\nu}$ one obtains the standard relativistic action 
supplemented with Lorentz symmetry breaking terms, proportional to 
inverse powers of $\kappa$.

The next case is described by commutator \r{1-1} with linear term in 
$\hat x_\mu$.
\be\label{1-3}
[\hat x_\mu, \hat x_\nu]=i \frac1\kappa \theta\b1_{\mu\nu}{}^\rho\hat 
x_\rho\,.
\ee
This is the main subject of our considerations. In order to obtain the 
translational invariance of the theory
\be\label{1-4}
\hat x_\mu \to \hat x_\mu+ \hat v_\mu
\ee
the basic algebra \r{1-3} imply that 
\bel{1-5}
[\hat v_\mu, \hat v_\nu] = i \frac1\kappa \theta\b1 
_{\mu\nu}{}^{\rho}\hat v_\rho
\ee
and
\bel{1-6}
[\hat x_\mu, \hat v_\nu] = 0\,.
\ee
The relations \r{1-5}-\r{1-6} imply that the addition formula \r{1-4} 
can be described as 
\bel{1-7}
\Delta(\hat x_\mu) = \hat x_\mu \tens 1 + 1 \tens x_\mu
\ee
where $\hat x_\mu = \hat x_\mu \tens 1$, $\hat v_\mu = 1\tens \hat 
x_\mu$. We see that the relations \r{1-3}-\r{1-7} describe the Hopf 
algebra of noncommutative translations, with primitive coproduct 
\r{1-7}.

An interesting question is to ask whether the noncommutative relations 
\r{1-3} can be extended do $D=4$ quantum Poincar\'e group, with 
generators
$(\hat x_\mu, \hat \Lambda_\mu{}^\nu)$ endowed with classical coproduct
\bel{1-8}
\ba{rcl}
\Delta (\hat \Lambda _\mu{}^\rho)&=& \hat \Lambda _\mu{}^\nu \hat\tens  
\Lambda _\nu{}^\rho \\[3mm]
\Delta (\hat x _\mu)&=& \hat x_\rho \tens  \Lambda ^\rho{}_\mu + 1 
\tens x_\mu\,. 
\ea
\ee
The general answer to this question is contained in the paper by
Podle\'s and Wo\-ro\-no\-wicz \cite{8}.
In the case of Lie-algebraic deformation \r{1-3} it appears \cite{9,10,11} 
that the most general choice permitting to extend \r{1-3} 
to Poincar\'e quantum group depends on a four-vector $a^\mu$ in the 
following way:
\bel{1-9}
\theta\b1_{\mu\nu}{}^\rho=a_\mu\delta_\nu{}^\rho-a_\nu\delta_\mu{}^\rho\,.
\ee
One can consider the following three cases:
\begin{enumerate}
\item[i)] If $a=(1,0,0,0)$ one obtains standard $\kappa$-deformation 
of Minkowski space, with quantum time coordinate $\hat t$ ($\hat x_0=c\hat t$)
\bel{1-10}
[\hat x_0, \hat x_i]= \frac i\kappa \hat x_i \,,\qquad
[\hat x_i, \hat x_j]= 0\,.
\ee
\item[ii)] If $a=\frac1{\sqrt2}(1,1,0,0)$ (light-like four-vector; $a_\mu a^\mu=0$)
one gets ($\hat x_\pm=\frac1{\sqrt2}(\hat x_0 \pm \hat x_3$); $r,s=1,2$)
\bel{1-11}
\ba{rclrcl}
[\hat x_+, \hat x_-]&=&\ds \frac i\kappa \hat x_- \,,\qquad&
[\hat x_+, \hat x_r]&=&\ds \frac i\kappa \hat x_r\,.\\[3mm]
[\hat x_-, \hat x_r]&=&0 \,,\qquad&
[\hat x_r, \hat x_s]&=& 0\,.
\ea
\ee
\item [iii)] If $a=(0,1,0,0)$ one obtains the tachyonic 
$\kappa$-deformation, with quantum third space coordinate 
\bel{1-12}
\ba{rclrcl}
[\hat x_3, \hat x_0]&=&\ds \frac i\kappa \hat x_0 \,,\qquad&
[\hat x_3, \hat x_r]&=&\ds \frac i\kappa \hat x_r\,,\\[3mm]
[\hat x_0, \hat x_r]&=& 0 \,,\qquad &
[\hat x_r, \hat x_s]&=& 0\,.
\ea
\ee
\end{enumerate}
The generalized $\kappa$-deformations, described by the choice \r{1-9} 
of noncommutative space-time was proposed firstly in \cite{9} (see also
\cite{10,11,12}).

In Sect. 2 we shall describe some properties of standard 
$\kappa$-deformed symmetries. We shall consider as basic 
characterization of these quantum symmetries the $\kappa$-deformed 
Poincar\'e algebra which is dual as a Hopf algebra (see e.g.\ \cite{13}) to
the $\kappa$-deformed 
Poincar\'e group. We shall notice here some new results concerning 
finite $\kappa$-deformed Lorentz transformations \cite{14}.

In Sect. 3 we shall discuss the light-like $\kappa$-deformation and its 
extension to $\kappa$-deformation od$d=4$ conformal symmetries. It 
appears that only the $a_\mu$-dependent $\kappa$-deformations with 
$a_\mu a^\mu=0$ are described by classical $r$-matrices satisfying the 
classical Yang-Baxter equation. It has been shown \cite{15} that  such $r$-matrix 
can be quantized by the extended Jordanian twist method \cite{16,17,18}.

In Sect. 4 we discuss some recent ideas related with the astrophysical 
applications of $\kappa$-deformed space-time.

\section{Standard $\kappa$-deformation}

Standard $\kappa$-deformation has been obtained firstly (see \cite{19}) by 
contraction of $q$-deformed anti-de-Sitter algebra. In such a 
formulation of the $\kappa$-deformed Poincar\'e algebra the
 nonrelativistic $E(3)$ algebra ($O(3)$ generators $M_i$ + 
space translation generators $P_i$) supplemented by $P_0$
remains classical. We call such a basis standard. Subsequently 
by suitable change of basis \cite{20} there was obtained 
$\kappa$-deformed Poincar\'e algebra in bicrossproduct basis:
\begin{enumerate} 
\item [a)] algebraic sector 
($M_{\mu\nu}=(M_i=\frac12\epsilon_{ijk}M_{jk}$, $N_i=M_{i0}$), 
$P_\mu=(P_i,P_0)$)
\bel{2-1}
\ba{rcl}
[M_{\mu\nu} ,M_{\rho\tau}]&=&i(\eta_{\mu\tau}M_{\nu\rho}
                             -\eta_{\mu\rho}M_{\nu\tau}
			     +\eta_{\nu\rho}M_{\mu\tau}
			     -\eta_{\nu\tau}M_{\mu\rho}) \\[2mm]
[M_i,P_j]&=&i\epsilon_{ijk}P_k\,, \qquad [M_i,P_0]\,=\,0\,,\\[2mm]
[N_i,P_j]&=& i \delta_{ij} [\frac \kappa 2 (1-e^{-\frac{2P_0}{\kappa}}) + 
\frac1{2\kappa} \vec P^2 + 
\frac1\kappa P_i P_j\,.\\[2mm]
[N_i,P_0]&=&iP_i\,,\\[2mm]
[P_\mu,P_\nu]&=&0\,.
\ea
\ee
\item [b)] coalgebraic sector
\bel{2-2}
\ba{rcl}
\Delta M_i &=& M_i \tens 1 + 1 \tens M_i \,,\\[2mm]
\Delta N_i &=& N_i \tens 1 + e^{-\frac{P_0}\kappa} \tens N_i
+\frac1\kappa\epsilon_{ijk}P_j\tens M_k \,,\\[2mm]
\Delta P_i &=& P_i \tens 1 + e^{-\frac{P_0}\kappa} \tens P_i \,,\\[2mm]
\Delta P_0 &=& P_0 \tens 1 + 1 \tens P_0 \,,\\[2mm]
\ea
\ee
\item[c)] antipodes
\bel{2-3}
\ba{rcl}
S(M_i)&=&-M_i\,,\\[2mm]
S(N_i)&=& -e^{\frac{P_0}\kappa}N_i+\frac1\kappa \epsilon_{ijk} 
e^{\frac{P_0}\kappa} P_j M_k\\[2mm]
S(P_i)&=&-e^{\frac{P_0}\kappa} P_i\,,\\[2mm]
S(P_0)&=& - P_0\,.
\ea
\ee
\end{enumerate}
One can show that the
Hopf algebra \r{2-1}-\r{2-3} is described infinitesimally by the 
following classical $r$-matrix 
\bel{2-3A}
r_1=\frac1\kappa (N_i\wedge P_i)
\ee
satisfying modified Yang-Baxter equation. 
In bicrossproduct basis we see that 
in comparison with classical Poincar\'e symmetries
only the boost transformations of 
three-momenta are deformed. Further in bicrossproduct basis one can 
introduce consistently the finite Lorentz transformations of 
four-momentum generators by standard formulae:
\bel{2-4}
P_\mu(\vec\beta) = e^{i\vec \beta\vec N} P_\mu  e^{-i\vec \beta\vec N}=
\sum_{n=0}^\infty \frac{1^n}{n!} [\vec \beta \vec N,  [\vec \beta \vec N, 
\ldots [\vec \beta \vec N, P_\mu]]\ldots]
\ee
The basis relation in bicrossproduct basis which justifies the formula
\r{2-3A}
as consistent with Hopf algebra structure is the following
equality for quantum adjoint operator (see \cite{21},  chapt 2d)
\bel{2-5}
\mathrm{ad}_{N_i} f(P_\mu) = N\b1_i f(P_\mu) S(N\b2_i) = [N_i,f(P_\mu)]\,,
\ee
where $\Delta N_i= N\b1_i\tens N\b2_i$ (no summation!) is given by \r{2-2}.
The calculation of the function $P_\mu(\vec \beta)$ for the choice of 
boost $\vec \beta=(0,0,\beta)$ has been recently performed explicitly 
in \cite{14}.

The deformation of $D=4$ Poincar\'e algebra can be removed if we 
introduce nonlinear realization of the four-momentum operators \cite{21,22,23} 
\footnote{We choose particular deformation map. The general case is 
discussed in \cite{23}.}
\bel{2-6}
\ba{rcl}
P\b0_i &=& e^{\frac{P_0}{2\kappa}}P_i\,,\\[2mm]
P\b0_0&=& \kappa \sinh \frac{P_0}{\kappa} +\frac1{2\kappa}\vec P^2\,.
\ea
\ee
The generators $(M_{\mu\nu},P\b0_\mu$) satisfy classical Poincar\'e 
algebra, with quite complicated coproduct \cite{12}. Because the Lorentz 
transformations of the four-momenta $P\b0_\mu$ are classical, it follows 
from the formulae \r{2-5}-\r{2-6} that the finite Lorentz 
transformations in bicrossproduct basis can be obtained
from classical Lorentz transformations by
deformation map, described by the inverse formulae to the 
relations \r{2-6}:
\bel{2-7}
\ba{rcl}
P_i&=& \kappa \left(P\b0_0+ \sqrt{P\b0_\mu 
{P\b0}^\mu+\kappa^2}\right)^{-1}\cdot P\b0_i\,,\\[2mm]
P_0&=& \kappa \ln\left(\frac{P\b0_0+ \sqrt{P\b0_\mu 
{P\b0}^\mu+\kappa^2}}{\kappa}\right)\,.
\ea\ee

In the classical Poincar\'e algebra basis ($M_{\mu\nu}$, $P\b0_\mu$) 
the whole effect 
of deformation is contained in the coproduct. In particular we stress 
that the coproduct of the energy operator $P\b0_0$ is nonprimitive 
(nonabelian).
We would like to recall that in \cite{12} an attempt was undertaken to 
interpret the noncocommutative coproduct for the energy operator $P\b0_0$ as 
describing geometric two-particle interactions.

\section{Light-like $\kappa$-deformation in standard basis}

The $\kappa$-deformation in standard basis leading to "quantum" light-cone 
direction $x_+=\frac1{\sqrt2}(x_0+x_3)$ (see (11-12)) was firstly obtained in \cite{24}.
 Further in \cite{9}
such a deformation was obtained in bicrossproduct basis as a special case of 
generalized $\kappa$-deformation, depending on arbitrary four-vector $a_\mu$.
The equivalence of two formulations has been shown in  \cite{25} 
(see also \cite{26}).

Let us introduce the light-cone basis ($a_\mu=\frac1{\sqrt2}(1,1,0,0)$, 
$\widetilde a_\mu=\frac1{\sqrt2}(1,-1,0,0)$;
$a_\mu a^\mu= \widetilde a_\mu \widetilde a^\mu =0$, $a_\mu \widetilde a^\mu=1$; $r=1,2$)
\bel{3-0}
\ba{rclrclrcl}
\widetilde P_0&=& \widetilde a^\mu P_\mu =P_0 +P_3 \,,\qquad&	
\widetilde P_3&=&  a^\mu P_\mu =P_0 -P_3 \,,\qquad& \widetilde P_r&=&P_r\,, \\[2mm]
\widetilde M_{03}&=&\widetilde a^\mu  a^\nu M_{\mu\nu} \,,& 
\widetilde M_{0r}&=&\widetilde a^\mu  M_{\mu r} \,,\\[2mm]
\widetilde M_{3r}&=& a^\mu  M_{\mu r} \,,&
\widetilde M_{12}&=& M_{12} \,.
\ea
\ee
The light-like $\kappa$-deformation of Poincar\'e algebra related by duality 
with noncommutative translation sector described by eq. \r{1-11} is given by 
the following Hopf algebra.
\begin{enumerate}
\item[a)] algebraic sector $\widetilde M_i=\frac12 \epsilon _{ijk}\widetilde M_{jk}$;
$\widetilde N_i = \widetilde M_{i0} $
Besides the classical Lorentz algebra relations and classical 
commutativity of four-momenta we obtain
\bel{3-1}
\ba{rcl}
[\widetilde M_i,\widetilde P_0]&=&0\,,\\[2mm]
[\widetilde M_i,\widetilde P_j]&=&i \kappa \epsilon_{ij} (1 - e^{-\frac{\widetilde P_0}{\kappa}}) 
+ i \epsilon_{ijk} \widetilde  P_k \,,\\[2mm]
[\widetilde N_i, \widetilde P_0]&=& 
i \kappa \delta_{i3} (1 - e^{-\frac{\widetilde P_0}{\kappa}}) + i (1-\delta_{i3}) 
\widetilde P_i\,,\\[2mm]
[\widetilde N_i, \widetilde P_k]&=& 
i \delta_{ik} \widetilde P_3 e^{-\frac{\widetilde P_0}{\kappa}}
+i \delta_{i3} \widetilde P_k (e^{-\frac{\widetilde P_0}{\kappa}}-1) 
+\frac{i}{2\kappa} \delta_{ik} (\widetilde P_1^2+\widetilde P_2^2)\\[2mm]
&&-\frac{i}{\kappa}(1-\delta_{i3}) \widetilde P_i \widetilde P_k \,,
\ea
\ee
\item[b)] coalgebraic sector and antipodes are provided by the relations 
\r{2-2}-\r{2-3} 
with the replacement $M'_i\to \widetilde M_i$, $N_i \to \widetilde N_i$ and $P_\mu\to
\widetilde P_\mu$.
\end{enumerate}

The above Hopf algebra is described by the following classical
$r$-matrix (see also \cite{27})
\bel{3-4}
r_2=\frac1\kappa (\widetilde N_i \wedge \widetilde P_i) =
\frac1{\kappa\sqrt2} [N_3\wedge (P_0-P_3) + (N_2-M_1)\wedge P_2+
(N_1+M_2)\wedge P_1]
\ee
which satisfies classical Yang-Baxter equation
\bel{3-5}
[[r_2,r_2]]\equiv 
 [r_2\b{12},r_2\b{13}]
+[r_2\b{12},r_2\b{23}]
+[r_2\b{13},r_2\b{23}] =0
\ee
The classical $r$-matrix \r{3-4} can be promoted to the $r$-matrix 
for $D=4$ conformal group, described by $SU(4)\simeq O(4,2)$ algebra.
It appears that the generators ($\widetilde N_i$,$\widetilde P_i$)
spanning classical $r$-matrix \r{3-4} can be described by the generators 
from the positive Borel subalgebra of $sl(4)$.
With suitable reality conditions
(see e.g. \cite{27}) we obtain
\bel{3-5A}
r_2=\frac1\kappa \left[\textstyle \frac12 (h_1+h_3)\wedge e_6 +e_1\wedge e_5 - e_3\wedge e_4\right]\,,
\ee
where the root generators of $sl(4)$ enter into the fundamental $4\times 4$ matrix 
representation as follows (see e.g. \cite{28})
\bel{3-6}
\pmatrix{
h_1 & e_1 & e_4 &e_6 \cr
e_{-1} & h_2 & e_2 & e_5 \cr
c_{-4} & e_{-2} & h_3 & e_3 \cr
e_{-6} & e_{-5} & e_{-3} &h_4 
}
\ee
Using results presented in \cite{16,17,18} one can calculate the generalized 
twist function, providing the full Hopf algebra structure of 
$\kappa$-deformed  $D=4$ conformal algebra \cite{15}. In fact, we have 
looked also on multiparameter solution of classical YB equation,  
spanned by the generators belonging to Weyl algebra 
(Poincar\'e generator + dilatations). We found 
for the classical $r$-matrices carrying mass dimension the 
corresponding quantum $\kappa$-deformed conformal algebras
described by suitable extended twist functions.

\section{Outlook}
Recently $\kappa$-deformations as an example of noncommutative geometry 
supplemented with quantum covariance group were tried as describing 
quantum gravity effects (see e.g. \cite{29,30}) and were
applied to astrophysical phenomena
(see e.g. \cite{31,32,33}). 
In commutative four-momentum framework the following two ingredients of 
$\kappa$-deformed geometry are used:
\begin{enumerate}
\item[i)] The deformation of mass-shell condition\footnote{In 
bicrossproduct basis one can consider two deformed 
mass-shell formulae related by the inversion of $\kappa$: 
$\kappa\to\kappa^{-1}$ 
(see \cite{34}, formula (2.34)).}
\bel{27} \textstyle
{\vec P}^2 e^{\pm\frac{P_0}\kappa} - (2 \kappa \sinh 
\frac{P_0}{\kappa})^2 = - M^2
\ee
\item[ii)] The nonsymmetric, nonabelian composition law for the four-momenta, described 
by the coproduct of momentum generators (see \r{2-2}).
\end{enumerate}
Both these elements of new theory lead at present to some conceptual problems. 
If we assume that the parameter $\kappa$ is universal (besides $\hbar$ 
and $c$ as the third fundamental constant) the straightforward
application of formula
\r{27} to macroscopic bodies leads to 
obvious contradiction with experimental results\footnote{This 
difficulty was pointed out to the author firstly by I.\ 
Bia\-{\l}y\-nic\-ki-Bi\-ru\-la.}. We see that the range of 
applicability of the formula \r{27} is still not understood what is related 
with some new notion of elementarity in microworld physics still to be 
discovered. The difficulty with consistent
description by $\kappa$-physics of macroworld phenomena
is also connected with the appearance of
 nonsymmetric, nonabelian addition law for the 
momenta. The problem how to incorporate modified momentum conservation law
into physical models
remains to be solved before we can apply $\kappa$-deformed 
kinematics e.g.\ to scattering processes 
and description of $\kappa$-deformed Feynman diagrams \cite{35}.
At present stage 
there were proposed some new ideas (see e.g. \cite{36}) but according 
to our estimate they are quite preliminary.

\section*{Acknowledgment}
Support by the KBN grant 5P03B05620 is acknowledged.


\end{document}